\documentclass[fleqn,onecolumn,showpacs,preprintnumbers,showkeys]{revtex4}
\usepackage{graphicx}
\usepackage{amssymb}
\usepackage{amsmath}
\usepackage{bm}
\usepackage[flushleft]{caption2}

\begin{document}
\setcounter{page}{1}
\title{Scalar invariants in gravitational and electromagnetic fields}
\author{Zihua Weng}
\email{xmuwzh@xmu.edu.cn.}
\affiliation{School of Physics and
Mechanical \& Electrical Engineering,
\\Xiamen University, Xiamen 361005, China}

\begin{abstract}
The paper discusses some scalar invariants in the gravitational
field and electromagnetic field by means of the characteristics of
the quaternions. When we emphasize some definitions of quaternion
physical quantities, the speed of light, mass density, energy
density, power density, charge density, and spin magnetic moment
density etc. will remain the same respectively in the gravitational
and electromagnetic fields under the coordinate transformation. The
results explain why there are some relationships among different
invariants in the gravitational and electromagnetic fields.
\end{abstract}



\maketitle

\section{INTRODUCTION}

In the gravitational field, there exist the conservation of mass,
conservation of energy \cite{bender}, mass continuity equation, and
invariable energy density. A. Lavoisier \cite{lavoisier} and H.
Cavendish etc. proposed the mass was invariable and conserved. There
are not only the conservation of mass but also the mass continuity
equation. G. W. Leibnitz, J. P. Joule \cite{joule}, and O. Heaviside
etc. believed the energy was invariable and conserved. A. Einstein
\cite{einstein} and M. Planck etc. found the mass-energy equivalence
between the mass and energy.

However, there exist still some puzzles about the above
conservations. Firstly, the conservation of mass is limited to the
case of weak gravitational strength. Is it effective still when the
gravitational field can not be neglected? Secondly, the conservation
of mass possesses other clear mathematical formulations in some
theories, such as the mass continuity equation. Whether the
conservation of energy has similar mathematical formulations in some
theories? Thirdly, the energy can be converted into the mass, and
vice versa. Are there other relations between these two
conservations?

The quaternion can be used to describe the property of the
electromagnetic field \cite{maxwell}. Similarly, the quaternion can
also be used to demonstrate the characteristics of gravitational
field. By means of the scalar invariant of quaternions, we find
there exist some relationships among these conservations and
invariants. With the characteristics of the quaternion \cite{adler},
we obtain that the speed of light, mass density, and energy density
are all invariants in the gravitational field \cite{chu}, under the
quaternion coordinate transformation \cite{weng}.

\section{Scalar invariants in gravitational field}

By means of the characteristics of the quaternion, we can obtain
some kinds of scalar invariants under the quaternion coordinate
transformation in the gravitational field. In the paper, each
quaternion definition of a physical quantity possesses one invariant
equation.

\subsection{Quaternion transformation}

In the quaternion space for the gravitational field, the basis
vector is $\mathbb{E} = (\emph{\textbf{i}}_0, \emph{\textbf{i}}_1,
\emph{\textbf{i}}_2, \emph{\textbf{i}}_3)$ with $\emph{\textbf{i}}_0
= 1$, and the quaternion quantity $\mathbb{D}(d_0, d_1, d_2, d_3)$
is defined as follows.
\begin{equation}
\mathbb{D} = d_0 + d_1 \emph{\textbf{i}}_1 + d_2 \emph{\textbf{i}}_2
+ d_3 \emph{\textbf{i}}_3
\end{equation}
where, the $d_0,~d_1,~d_2$, and $d_3$ are all real.

When we transform one form of the coordinate system into another,
the physical quantity $\mathbb{D}$ will be transformed into the
quaternion physical quantity $\mathbb{D}' (d'_0 , d'_1 , d'_2 ,
d'_3)$.
\begin{equation}
\mathbb{D}' = \mathbb{K}^* \circ \mathbb{D} \circ \mathbb{K}
\end{equation}
where, the $\circ$ denotes the quaternion multiplication; the $*$
indicates the conjugate of the quaternion; the $\mathbb{K}$ is the
quaternion, and $\mathbb{K}^* \circ \mathbb{K} = 1$ .

The scalar part $d_0$ satisfies the following equation.
\begin{eqnarray}
d_0 = d'_0
\end{eqnarray}

In the above equation, both sides' scalar parts are one and the same
during the quaternion coordinate system is transforming.

\subsection{Radius vector}

In the quaternion space, the coordinates are $r_0 , r_1 , r_2$, and
$r_3$. The radius vector is
\begin{eqnarray}
\mathbb{R} = r_0 + \Sigma (r_j \emph{\textbf{i}}_j)
\end{eqnarray}
where, the scalar $r_0 = v_0 t$; $v_0$ is the speed of light beam,
and $t$ denotes the time; $j = 1, 2, 3$; $i = 0, 1, 2, 3 $.

When the coordinate system is transformed into the other, we have
the quaternion radius vector $\mathbb{R}' (r'_0 , r'_1 , r'_2 , r'_3
)$, with the $r'_0 = v'_0 t'$ from Eq.(2).

From Eqs.(3) and (4), we obtain
\begin{eqnarray}
r_0 = r'_0
\end{eqnarray}

In the some special cases, we may substitute the quaternion physical
quantity $\mathbb{Z} (z_0, z_1, z_2, z_3)$ for the radius vector
$\mathbb{R} (r_0, r_1, r_2, r_3)$. The former quantity is defined as
\begin{equation}
\mathbb{Z} = \mathbb{R} \circ \mathbb{R}
\end{equation}
where, $z_0 = (r_0)^2 - (r_1)^2 - (r_2)^2 - (r_3)^2$ .

When we transform one form of the coordinate system into another,
the physical quantity $\mathbb{Z}$ is transformed into the
quaternion physical quantity $\mathbb{Z}' (z'_0, z'_1, z'_2, z'_3)$
from Eq.(2).

From Eqs.(3) and (6), we gain
\begin{eqnarray}
(r_0)^2 - \Sigma (r_j)^2 = (r'_0)^2 - \Sigma (r'_j)^2
\end{eqnarray}

The above means clearly that we may obtain different invariants from
differen definitions of physical quantities, by means of the
characteristics of quaternion coordinate transformation in the
gravitational field.

\begin{table}[t]
\caption{\label{tab:table1}The quaternion multiplication table.}
\begin{ruledtabular}
\begin{tabular}{ccccc}
$ $ & $1$ & $\emph{\textbf{i}}_1$  & $\emph{\textbf{i}}_2$ &
$\emph{\textbf{i}}_3$  \\
\hline $1$ & $1$ & $\emph{\textbf{i}}_1$  & $\emph{\textbf{i}}_2$ &
$\emph{\textbf{i}}_3$  \\
$\emph{\textbf{i}}_1$ & $\emph{\textbf{i}}_1$ & $-1$ &
$\emph{\textbf{i}}_3$  & $-\emph{\textbf{i}}_2$ \\
$\emph{\textbf{i}}_2$ & $\emph{\textbf{i}}_2$ &
$-\emph{\textbf{i}}_3$ & $-1$ & $\emph{\textbf{i}}_1$ \\
$\emph{\textbf{i}}_3$ & $\emph{\textbf{i}}_3$ &
$\emph{\textbf{i}}_2$ & $-\emph{\textbf{i}}_1$ & $-1$
\end{tabular}
\end{ruledtabular}
\end{table}

\subsection{Invariable speed of light}

The quaternion velocity $\mathbb{V} (v_0, v_1, v_2, v_3)$ is
\begin{eqnarray}
\mathbb{V} = v_0 + \Sigma (v_j \emph{\textbf{i}}_j)
\end{eqnarray}

When the coordinate system is transformed into the other, we have
the quaternion velocity $\mathbb{V}' (v'_0, v'_1, v'_2, v'_3 )$ from
Eq.(2).

From Eq.(3) and the definition of velocity, we obtain the invariable
speed of light. That is
\begin{eqnarray}
v_0 = v'_0
\end{eqnarray}

Choosing the definition combination of the quaternion velocity and
radius vector, we find the invariants, Eqs.(5) and (9), and gain
Galilean transformation.
\begin{eqnarray}
r_0 = r'_0~,~v_0 = v'_0~.
\end{eqnarray}

In some cases, we obtain Lorentz transformation from the invariant
equations, Eqs.(7) and (9).
\begin{eqnarray}
(r_0)^2 - \Sigma (r_j)^2 = (r'_0)^2 - \Sigma (r'_j)^2~,~v_0 = v'_0~.
\end{eqnarray}

The above states that we can obtain many kinds of coordinate
transformations from the different definition combinations of
physical quantities, including Galilean transformation and Lorentz
transformation, etc.

\subsection{Potential and strength}

In the quaternion space, the gravitational potential is,
\begin{eqnarray}
\mathbb{A} = a_0 + \Sigma (a_j \emph{\textbf{i}}_j)
\end{eqnarray}
and the gravitational strength $\mathbb{B}(b_0, b_1, b_2, b_3)$ as
follows.
\begin{eqnarray}
\mathbb{B} = \lozenge \circ \mathbb{A}
\end{eqnarray}
where, $\partial_i = \partial/\partial r_i$; $ \lozenge = \Sigma
(\emph{\textbf{i}}_i \partial_i)$; $b_0 = \partial_0 a_0 - \Sigma
(\partial_j a_j)$.

In the above equation, we choose the following gauge condition to
simplify succeeding calculation,
\begin{eqnarray}
b_0 = 0
\end{eqnarray}
and then the gravitational strength is $\mathbb{B} = \textbf{g}/v_0
+ \textbf{b}$ .
\begin{eqnarray}
\textbf{g}/v_0 = && \partial_0 \emph{\textbf{a}} + \nabla a_0
\\
\textbf{b} = && \nabla \times \emph{\textbf{a}}
\end{eqnarray}
where, $\nabla = \Sigma (\emph{\textbf{i}}_j \partial_j)$;
$\emph{\textbf{a}} = \Sigma (a_j \emph{\textbf{i}}_j)$;
$\emph{\textbf{b}} =  \Sigma (b_j \emph{\textbf{i}}_j)$; there exist
$\textbf{a} = 0$ and $\textbf{b} = 0$ in Newtonian gravity.

When the coordinate system is transformed into the other, we have
the quaternion potential $\mathbb{A}' (a'_0, a'_1, a'_2, a'_3 )$ and
strength $\mathbb{B}' (b'_0, b'_1, b'_2, b'_3 )$ respectively from
Eq.(2).

From Eq.(3) and the definition of potential, we obtain the
invariable scalar potential,
\begin{eqnarray}
a_0 = a'_0
\end{eqnarray}
and the gauge from the definition of strength as follows.
\begin{eqnarray}
b_0 = b'_0
\end{eqnarray}

By means of the quaternion characteristics, the above means that the
scalar potential of gravitational field is the invariant under the
coordinate transformations in Eq.(2). And so is that of
gravitational strength.

\subsection{Conservation of mass}

There exist two kinds of linear momenta, one is of the particle, the
other is of the gravitational field. For the sake of describing
various linear momenta by one single definition, the source and the
linear momentum both will be extended in the gravitational field.

The gravitational source $\mathbb{S}$ covers the linear momentum
density $\mathbb{S}_g$ and an extra part $ \mathbb{B}^* \circ
\mathbb{B}/(v_0 \mu_g^g) $.
\begin{eqnarray}
\mu \mathbb{S} = - ( \mathbb{B}/v_0 + \lozenge)^* \circ \mathbb{B} =
\mu_g^g \mathbb{S}_g - \mathbb{B}^* \circ \mathbb{B}/v_0
\end{eqnarray}
where, $\mathbb{B}^* \circ \mathbb{B}/(2\mu_g^g)$ is the energy
density of gravitational field, and is similar to that of the
electromagnetic field; $\mu$ and $\mu_g^g$ are the constants;
$\mu_g^g = 4 \pi G / v_0^2$ , and $G$ is the gravitational constant.

The linear momentum density $\mathbb{P}(p_0, p_1, p_2, p_3)$ is the
extension of $\mathbb{S}_g = m \mathbb{V}$.
\begin{eqnarray}
\mathbb{P} = \mu \mathbb{S} / \mu_g^g
\end{eqnarray}
where, $p_0 = \widehat{m} v_0$ , $p_j = m v_j \emph{\textbf{i}}_j$ ;
$\widehat{m} = m + \triangle m $; $m$ is the inertial mass density,
and $\widehat{m}$ is the gravitational mass density; $\triangle m =
- \mathbb{B}^* \circ \mathbb{B}/(v_0^2 \mu_g^g)$.

When the quaternion coordinate system rotates, we have the
quaternion linear momentum density $\mathbb{P}'(p'_0, p'_1, p'_2,
p'_3)$ from Eqs.(3) and (20).
\begin{eqnarray}
\widehat{m}v_0 = \widehat{m}' v'_0
\end{eqnarray}

By Eqs.(9) and (21), the gravitational mass density $\widehat{m}$
remains unchanged in the gravitational field.
\begin{eqnarray}
\widehat{m} = \widehat{m}'
\end{eqnarray}

The above is the conservation of mass. When $b_j = 0$ and $\triangle
m = 0$, the above can be reduced as follows.
\begin{eqnarray}
m = m'
\end{eqnarray}

The above means that if we emphasize the definitions of the velocity
and linear momentum, the gravitational mass density will remain the
same under the coordinate transformation in Eq.(2) in the quaternion
space.

\subsection{Mass continuity equation}

We choose the following definition of applied force, to cover
various forces in the gravitational field. The applied force density
$\mathbb{F}(f_0, f_1, f_2, f_3)$ is defined from the linear momentum
density $\mathbb{P}$ .
\begin{eqnarray}
\mathbb{F} = v_0 (\mathbb{B}/v_0 + \lozenge )^* \circ \mathbb{P}
\end{eqnarray}
where, the applied force density includes the inertial force density
and the gravitational force density, etc. While, the scalar $f_0 =
\partial p_0 / \partial t + v_0 \Sigma (  \partial p_j / \partial
r_j ) + \Sigma ( b_j p_j ) $ .

When the coordinate system rotates, we have the quaternion applied
force density $\mathbb{F}' (f'_0 , f'_1 , f'_2 , f'_3 )$ from
Eqs.(2), (3) and (24). And then, we have
\begin{eqnarray}
f_0 = f'_0
\end{eqnarray}

In the equilibrium state, there exists $f'_0 = 0$, we have the mass
continuity equation by Eqs.(9) and (25).
\begin{eqnarray}
\partial \widehat{m} / \partial t + \Sigma (  \partial p_j /
\partial r_j ) + \Sigma ( b_j p_j ) / v_0 = 0
\end{eqnarray}
further, if the strength $b_j = 0$, the above is reduced to the
differential form of mass continuity equation.
\begin{eqnarray}
\partial m / \partial t + \Sigma ( \partial p_j /
\partial r_j ) = 0
\end{eqnarray}

The above states the gravitational strength $\mathbb{B}$ has a small
influence on the mass continuity equation, although the term $\Sigma
( b_j p_j ) / v_0$ and the extra mass density $\triangle m$ both are
very tiny. And then the mass continuity equation is the invariant
under the quaternion transformation in Eq.(2), if we choose the
definition combination of the velocity and applied force in the
gravitational field.

\subsection{Angular momentum}

In order to incorporate various energies with the single definition,
the angular momentum and the energy both will be extended in the
gravitational field.

The angular momentum density $\mathbb{L}$ is defined from the linear
momentum density $\mathbb{P}$, radius vector $\mathbb{R}$, and
$\mathbb{X}$.
\begin{eqnarray}
\mathbb{L} = (\mathbb{R} + k_{rx} \mathbb{X} ) \circ \mathbb{P}
\end{eqnarray}
where, the $\mathbb{X} = \Sigma (x_i \emph{\textbf{i}}_i)$ is the
quaternion; the $k_{rx}$ is the coefficient; the angular momentum
density $\mathbb{L}$ includes the orbital angular momentum density
and spin angular momentum density in the gravitational field.

The angular momentum density is rewritten as,
\begin{eqnarray}
\mathbb{L} = l_0 + \Sigma (l_j \emph{\textbf{i}}_j )
\end{eqnarray}
where, $l_0 = (r_0 + k_{rx} x_0) p_0 - \Sigma \left\{ (r_j + k_{rx}
x_j) p_j \right\}$.

The scalar part $l_0$ is presumed to be the density of spin angular
momentum \cite{uhlenbeck} in the gravitational field. As a part of
the $\mathbb{L}$, the $l_0$ is one kind of angular momentum, and is
the scalar along the $\emph{\textbf{i}}_0$ direction. When the
$\mathbb{X}$ is neglected and the gravity is weak, the scalar $l_0$
will be reduced to $l_0 = r_0 p_0 - \Sigma (r_j p_j )$.

When the coordinate system rotates, we have the quaternion angular
momentum density $\mathbb{L}' (l'_0 , l'_1 , l'_2 , l'_3 )$ from
Eq.(2). Under the coordinate transformation, the spin angular
momentum density remains unchanged by Eq.(3).
\begin{eqnarray}
l_0 = l'_0
\end{eqnarray}

The above means that the gravitational strength has an influence on
the orbital angular momentum and the spin angular momentum, etc.
Choosing the definition of angular momentum, the spin angular
momentum will be invariable in the gravitational field under the
quaternion transformation in Eq.(2).

\begin{table}[b]
\caption{\label{tab:table1}The definitions and their scalar
invariants of the gravitational field in the quaternion space.}
\begin{ruledtabular}
\begin{tabular}{llll}
$definition$                             & $invariant $       &    $ meaning $                  & $theorem$          \\
\hline
$\mathbb{R}$                             & $r_0 = r'_0$       &    $invariable~scalar$          & $-$                \\
$\mathbb{V}$                             & $v_0 = v'_0$       &    $invariable~speed~of~light$  & $-$                \\
$\mathbb{X}$                             & $x_0 = x'_0$       &    $invariable~scalar$          & $-$                \\
$\mathbb{A}$                             & $a_0 = a'_0$       &    $invariable~gravitational~scalar~potential$ & $-$ \\
$\mathbb{B}$                             & $b_0 = b'_0$       &    $invariable~gravitational~gauge$            & $-$ \\
$\mathbb{P}$                             & $p_0 = p'_0$       &    $conservation~of~mass$       & $-$                \\
$\mathbb{F}$                             & $f_0 = f'_0$       &    $mass~continuity~equation$   & $linear~momentum$  \\
$\mathbb{L}$                             & $l_0 = l'_0$       &    $invariable~spin~angular~momentum~density$  & $-$ \\
$\mathbb{W}$                             & $w_0 = w'_0$       &    $conservation~of~energy$     & $angular~momentum$ \\
$\mathbb{N}$                             & $n_0 = n'_0$       &    $energy~continuity~equation$ & $torque$           \\
\end{tabular}
\end{ruledtabular}
\end{table}

\subsection{Conservation of energy}

We choose the quaternion definition of total energy to include
various energies in the gravitational field. In the quaternion
space, the total energy density $\mathbb{W}$ is defined from the
angular momentum density $\mathbb{L}$ .
\begin{eqnarray}
\mathbb{W}  = v_0 ( \mathbb{B}/v_0 + \lozenge) \circ \mathbb{L}
\end{eqnarray}
where, the total energy includes the potential energy, the kinetic
energy, the torque, and the work, etc.

The total energy density is,
\begin{eqnarray}
\mathbb{W} = w_0 + \Sigma (w_j \emph{\textbf{i}}_j )
\end{eqnarray}
where, the scalar $w_0 = \emph{\textbf{b}} \cdot \emph{\textbf{l}} +
v_0 \partial_0 l_0 + v_0 \nabla \cdot \emph{\textbf{l}}$ is the
energy density; $\emph{\textbf{w}} = \Sigma (w_j \emph{\textbf{i}}_j
)$; $\emph{\textbf{l}} = \Sigma (l_j \emph{\textbf{i}}_j)$.

When the coordinate system rotates, we have the quaternion total
energy density $\mathbb{W}' (w'_0 , w'_1 , w'_2 , w'_3 )$ from
Eq.(2). And under the coordinate transformation, the conservation of
energy will be obtained from the above and Eq.(3).
\begin{eqnarray}
w_0 = w'_0
\end{eqnarray}

In some special cases, there exists $w'_0 = 0$, we obtain the spin
continuity equation by Eq.(33).
\begin{eqnarray}
\partial l_0 / \partial r_0 + \nabla \cdot \emph{\textbf{l}}
+ \emph{\textbf{b}} \cdot \emph{\textbf{l}} / v_0 = 0
\end{eqnarray}

If the last term is neglected, the above is reduced to
\begin{eqnarray}
\partial l_0 / \partial r_0 + \nabla \cdot \emph{\textbf{l}} = 0
\end{eqnarray}
further, if the last term is equal to zero, we have
\begin{eqnarray}
\partial l_0 / \partial t = 0
\end{eqnarray}
where, when the time $t$ is only the independent variable, the
$\partial l_0 / \partial t$ will become the $d l_0 / d t$ .

The above means that the energy density $w_0$ is the invariant under
the quaternion transformation in Eq.(2) in the gravitational field.
The spin continuity equation is also the invariable in the
gravitational field, although the gravitational strength has an
influence on the spin angular momentum density.

\subsection{Energy continuity equation}

The external power can be described by the quaternion in the
gravitational field. The external power density $\mathbb{N}$ is
defined from the total energy density $\mathbb{W}$ .
\begin{eqnarray}
\mathbb{N}  = v_0 ( \mathbb{B}/v_0 + \lozenge)^* \circ \mathbb{W}
\end{eqnarray}
where, the external power density $\mathbb{N}$ includes the power
density etc. in the gravitational field.

The external power density is also,
\begin{eqnarray}
\mathbb{N} = n_0 + \Sigma (n_j \emph{\textbf{i}}_j )
\end{eqnarray}
where, the scalar $n_0 = \emph{\textbf{b}}^* \cdot \emph{\textbf{w}}
+ v_0 \partial_0 w_0 + v_0 \nabla^* \cdot \emph{\textbf{w}} $ is the
power density.

When the quaternion coordinate system rotates, we have the
quaternion external power density $\mathbb{N}' (n'_0 , n'_1 , n'_2 ,
n'_3 )$ from Eq.(2). Under the coordinate transformation, the power
density will remain unchanged by Eq.(3).
\begin{eqnarray}
n_0 = n'_0
\end{eqnarray}

In case of $n'_0 = 0$, we obtain the energy continuity equation from
the above.
\begin{eqnarray}
\partial w_0 / \partial r_0 + \nabla^* \cdot \emph{\textbf{w}}
+ \emph{\textbf{b}}^* \cdot \emph{\textbf{w}} / v_0 = 0
\end{eqnarray}

If the last term is neglected, the above is reduced to
\begin{eqnarray}
\partial w_0 / \partial r_0 + \nabla^* \cdot \emph{\textbf{w}} = 0
\end{eqnarray}
further, if the last term is equal to zero, we have
\begin{eqnarray}
\partial w_0 / \partial t = 0
\end{eqnarray}
where, when the time $t$ is only the independent variable, the
$\partial w_0 / \partial t$ will become the $d w_0 / d t$ .

The above means that the gravitational strength $\emph{\textbf{b}}$
and the torque density $\emph{\textbf{w}}$ have the influence on the
energy continuity equation. Meanwhile, the power density $n_0$ is
the invariant under the quaternion transformation in the
gravitational field. The energy continuity equation is also the
invariant, when we choose the definition combination of the
velocity, total energy, and external power.

In the quaternion spaces, the deductive conclusions about the
conservations and their relationships depend on the definition
combinations of physical quantities. It leads us to understand the
conservations more deeply. By means of the definition combination of
the radius vector and the velocity, we obtain the conclusions of the
invariable speed of light, Galilean transformation, and Lorentz
transformation etc. in the gravitational field. Alike, choosing the
definition combination of the linear momentum, the operator
$\lozenge$, and the invariable speed of light, we have the results
of the conservation of mass and the mass continuity equation.
Moreover, depending on the definition combination of the angular
momentum, the invariable speed of light, and the operator
$\lozenge$, we obtain the inference of the invariable spin density,
conservation of energy, and energy continuity equation, etc.

\section{Mechanical invariants in electromagnetic field}

In the electromagnetic field, some scalar invariants will be
extended from the gravitational field. They are called the
mechanical invariants, which include the conservation of mass,
conservation of spin, conservation of power, and conservation of
energy \cite{young}. In the electromagnetic field, there exist still
some puzzles about the conservation laws. Firstly, those
conservation laws are only limited to the case of weak
electromagnetic strength. Are they still effective when the
electromagnetic strength can not be neglected? Secondly, there exist
either the conservation of mass or the mass continuity equation. Are
there some similar continuity equations for other conservation laws?
Thirdly, the energy can be converted into the mass, and vice versa.
Are there other relations among these conservation laws?

The octonion \cite{cayley} can be used to describe the property of
the electromagnetic field and gravitational field, although these
two kinds of fields are quite different. By means of the scalar
invariant of the octonion, we find that the strength of the
electromagnetic field and gravitational field \cite{newton} have the
influence on the conservation laws.

With the character of the octonion, we obtain the mass density and
the energy density are invariants in the case for coexistence of
electromagnetic field and gravitational field, under the octonion
transformation, although a few definitions of physical quantities
will be extended in the gravitational field and electromagnetic
field.

\subsection{Octonion transformation}

In the quaternion space, the basis vector for the gravitational
field is $\mathbb{E}_g$ = ($1$, $\emph{\textbf{i}}_1$,
$\emph{\textbf{i}}_2$, $\emph{\textbf{i}}_3$), and that for the
electromagnetic field is $\mathbb{E}_e$ = ($\emph{\textbf{I}}_0$,
$\emph{\textbf{I}}_1$, $\emph{\textbf{I}}_2$,
$\emph{\textbf{I}}_3$). The $\mathbb{E}_e$ is independent of the
$\mathbb{E}_g$, with $\mathbb{E}_e$ = $\mathbb{E}_g \circ
\emph{\textbf{I}}_0$ . The basis vectors $\mathbb{E}_g$ and
$\mathbb{E}_e$ can be combined together to become the basis vector
$\mathbb{E}$ of octonion space.
\begin{eqnarray}
\mathbb{E} = (1, \emph{\textbf{i}}_1, \emph{\textbf{i}}_2,
\emph{\textbf{i}}_3, \emph{\textbf{I}}_0, \emph{\textbf{I}}_1,
\emph{\textbf{I}}_2, \emph{\textbf{I}}_3)
\end{eqnarray}

The octonion quantity $\mathbb{D} (d_0, d_1, d_2, d_3, D_0, D_1,
D_2, D_3 )$ is defined as follows.
\begin{eqnarray}
\mathbb{D} = d_0 + \Sigma (d_j \emph{\textbf{i}}_j) + \Sigma (D_i
\emph{\textbf{I}}_i)
\end{eqnarray}
where, $d_i$ and $D_i$ are all real; $i = 0, 1, 2, 3$; $j = 1, 2,
3$.

When the coordinate system is transformed into the other one, the
physical quantity $\mathbb{D}$ will be become the octonion physical
quantity $\mathbb{D}' (d'_0 , d'_1 , d'_2 , d'_3 , D'_0 , D'_1 ,
D'_2 , D'_3 )$ .
\begin{equation}
\mathbb{D}' = \mathbb{K}^* \circ \mathbb{D} \circ \mathbb{K}
\end{equation}
where, $\mathbb{K}$ is the octonion, and $\mathbb{K}^* \circ
\mathbb{K} = 1$; $*$ denotes the conjugate of octonion; $\circ$ is
the octonion multiplication.

Therefore
\begin{eqnarray}
d_0 = d'_0
\end{eqnarray}

In the above equation, the scalar part is one and the same during
the octonion coordinates are transforming. Some scalar invariants of
electromagnetic field will be obtained from the characteristics of
octonions.

\begin{table}[t]
\caption{\label{tab:table1}The octonion multiplication table.}
\begin{ruledtabular}
\begin{tabular}{ccccccccc}
$ $ & $1$ & $\emph{\textbf{i}}_1$  & $\emph{\textbf{i}}_2$ &
$\emph{\textbf{i}}_3$  & $\emph{\textbf{I}}_0$  &
$\emph{\textbf{I}}_1$
& $\emph{\textbf{I}}_2$  & $\emph{\textbf{I}}_3$  \\
\hline $1$ & $1$ & $\emph{\textbf{i}}_1$  & $\emph{\textbf{i}}_2$ &
$\emph{\textbf{i}}_3$  & $\emph{\textbf{I}}_0$  &
$\emph{\textbf{I}}_1$
& $\emph{\textbf{I}}_2$  & $\emph{\textbf{I}}_3$  \\
$\emph{\textbf{i}}_1$ & $\emph{\textbf{i}}_1$ & $-1$ &
$\emph{\textbf{i}}_3$  & $-\emph{\textbf{i}}_2$ &
$\emph{\textbf{I}}_1$
& $-\emph{\textbf{I}}_0$ & $-\emph{\textbf{I}}_3$ & $\emph{\textbf{I}}_2$  \\
$\emph{\textbf{i}}_2$ & $\emph{\textbf{i}}_2$ &
$-\emph{\textbf{i}}_3$ & $-1$ & $\emph{\textbf{i}}_1$  &
$\emph{\textbf{I}}_2$  & $\emph{\textbf{I}}_3$
& $-\emph{\textbf{I}}_0$ & $-\emph{\textbf{I}}_1$ \\
$\emph{\textbf{i}}_3$ & $\emph{\textbf{i}}_3$ &
$\emph{\textbf{i}}_2$ & $-\emph{\textbf{i}}_1$ & $-1$ &
$\emph{\textbf{I}}_3$  & $-\emph{\textbf{I}}_2$
& $\emph{\textbf{I}}_1$  & $-\emph{\textbf{I}}_0$ \\
$\emph{\textbf{I}}_0$ & $\emph{\textbf{I}}_0$ &
$-\emph{\textbf{I}}_1$ & $-\emph{\textbf{I}}_2$ &
$-\emph{\textbf{I}}_3$ & $-1$ & $\emph{\textbf{i}}_1$
& $\emph{\textbf{i}}_2$  & $\emph{\textbf{i}}_3$  \\
$\emph{\textbf{I}}_1$ & $\emph{\textbf{I}}_1$ &
$\emph{\textbf{I}}_0$ & $-\emph{\textbf{I}}_3$ &
$\emph{\textbf{I}}_2$  & $-\emph{\textbf{i}}_1$
& $-1$ & $-\emph{\textbf{i}}_3$ & $\emph{\textbf{i}}_2$  \\
$\emph{\textbf{I}}_2$ & $\emph{\textbf{I}}_2$ &
$\emph{\textbf{I}}_3$ & $\emph{\textbf{I}}_0$  &
$-\emph{\textbf{I}}_1$ & $-\emph{\textbf{i}}_2$
& $\emph{\textbf{i}}_3$  & $-1$ & $-\emph{\textbf{i}}_1$ \\
$\emph{\textbf{I}}_3$ & $\emph{\textbf{I}}_3$ &
$-\emph{\textbf{I}}_2$ & $\emph{\textbf{I}}_1$  &
$\emph{\textbf{I}}_0$  & $-\emph{\textbf{i}}_3$
& $-\emph{\textbf{i}}_2$ & $\emph{\textbf{i}}_1$  & $-1$ \\
\end{tabular}
\end{ruledtabular}
\end{table}

\subsection{Radius vector}

The radius vector is $\mathbb{R}_g$ = ($r_0$, $r_1$, $r_2$, $r_3$)
in the quaternion space for gravitational field. For the
electromagnetic field, the radius vector is $\mathbb{R}_e$ = ($R_0$,
$R_1$, $R_2$, $R_3$). Their combination is $\mathbb{R} (r_0 , r_1 ,
r_2 , r_3 , R_0 , R_1 , R_2 , R_3 )$.
\begin{eqnarray}
\mathbb{R} = r_0 + \Sigma (r_j \emph{\textbf{i}}_j) + \Sigma (R_i
\emph{\textbf{I}}_i)
\end{eqnarray}
where, $r_0 = v_0 t$, $R_0 = V_0 T$. $v_0$ is the speed of light,
$t$ denotes the time; and $V_0$ is the speed of light-like, $T$ is a
time-like quantity.

When the coordinates are rotated, we have the octonion radius vector
$\mathbb{R}' (r'_0, r'_1, r'_2, r'_3, R'_0, R'_1, R'_2, R'_3 )$ from
Eq.(45). And then, we have following consequence from Eqs.(46) and
(47).
\begin{eqnarray}
r_0 = r'_0
\end{eqnarray}

Sometimes, the radius vector $\mathbb{R}$ can be replaced by the
physical quantity $\mathbb{Z} (z_0, z_1, z_2, z_3, Z_0, Z_1, Z_2,
Z_3)$. And
\begin{eqnarray}
\mathbb{Z} = \mathbb{R} \circ \mathbb{R} = z_0 + \Sigma (z_j
\emph{\textbf{i}}_j) + \Sigma (Z_i \emph{\textbf{I}}_i)
\end{eqnarray}

By Eqs.(46) and (49), we have
\begin{eqnarray}
(r_0)^2 - \Sigma (r_j)^2 - \Sigma (R_i)^2 = (r'_0)^2 - \Sigma
(r'_j)^2 - \Sigma (R'_i)^2
\end{eqnarray}

The above represents that the spacetime interval $z_0$ remains
unchanged when the coordinate system rotates in the octonion space.

\subsection{Speed of light}

In the quaternion space for the gravitational field, the velocity is
$\mathbb{V}_g$ = ($v_0$, $v_1$, $v_2$, $v_3$). For the
electromagnetic field, the velocity is $\mathbb{V}_e$ = ($V_0$,
$V_1$, $V_2$, $V_3$).

They can be combined together to become the octonion velocity
$\mathbb{V} (v_0 , v_1 , v_2 , v_3 , V_0 , V_1 , V_2 , V_3 )$.
\begin{eqnarray}
\mathbb{V} = v_0 [ \lozenge \circ \mathbb{R} - \nabla \cdot \left\{
\Sigma ( r_j \emph{\textbf{i}}_j ) \right\} ] = v_0 + \Sigma (v_j
\emph{\textbf{i}}_j) + \Sigma (V_i \emph{\textbf{I}}_i)
\end{eqnarray}

When the coordinates are rotated, we have the octonion velocity
$\mathbb{V}' (v'_0 , v'_1 , v'_2 , v'_3 , V'_0 , V'_1 , V'_2 , V'_3
)$ from Eq.(45). And then, we have the invariable speed of light by
Eq.(46).
\begin{eqnarray}
v_0 = v'_0
\end{eqnarray}
and the Galilean transformation as follows.
\begin{eqnarray}
t_0 = && t'_0
\\
\Sigma (r_j)^2 + \Sigma (R_i)^2 = && \Sigma (r'_j)^2 + \Sigma
(R'_i)^2
\end{eqnarray}

The above means that if we emphasize especially the important of
radius vector Eq.(47) and velocity Eq.(51), we obtain Galilean
transformation from Eqs.(48) and (52).

In the same way, if we choose the definitions of radius vector
Eq.(49) and velocity Eq.(51), we obtain the Lorentz transformation
\cite{lorentz} from Eqs.(50) and (52).

In some special cases, the $\Sigma (R_i \emph{\textbf{I}}_i)$ do not
take part the octonion coordinate transformation, we have the result
$\Sigma (R_i)^2 = \Sigma (R'_i)^2$ in Eqs.(50) and (54).

\subsection{Potential and strength}

The gravitational potential is $\mathbb{A}_g = (a_0 , a_1 , a_2 ,
a_3)$, and the electromagnetic potential is $\mathbb{A}_e = (A_0 ,
A_1 , A_2 , A_3)$. And the gravitational potential and the
electromagnetic potential constitute the potential $\mathbb{A}$ .
\begin{eqnarray}
\mathbb{A} = v_0 \lozenge \circ \mathbb{X} = \mathbb{A}_g + k_{eg}
\mathbb{A}_e
\end{eqnarray}
where, $k_{eg}$ is the coefficient.

When the coordinates are rotated, we obtain the octonion potential
$\mathbb{A}' (a'_0, a'_1, a'_2, a'_3, A'_0, A'_1, A'_2, A'_3)$ from
Eq.(45). Therefore, we have the invariable scalar potential of
gravitational field by Eq.(46).
\begin{eqnarray}
a_0 = a'_0
\end{eqnarray}

The octonion strength $\mathbb{B}(b_0, b_1, b_2, b_3, B_0, B_1, B_2,
B_3)$ consists of the gravitational strength $\mathbb{B}_g$ and
electromagnetic strength $\mathbb{B}_e$.
\begin{eqnarray}
\mathbb{B} = \lozenge \circ \mathbb{A} = \mathbb{B}_g + k_{eg}
\mathbb{B}_e
\end{eqnarray}
where, the operator $\lozenge = \partial_0 + \emph{\textbf{i}}_1
\partial_1 + \emph{\textbf{i}}_2 \partial_2 + \emph{\textbf{i}}_3
\partial_3$ .

In the above equation, we choose the following gauge conditions to
simplify succeeding calculation.
\begin{eqnarray}
b_0 = \partial_0 a_0 + \nabla \cdot \textbf{a} = 0~,~ B_0 =
\partial_0 A_0 + \nabla \cdot \textbf{A} = 0~.
\nonumber
\end{eqnarray}
where, $\textbf{a} = \Sigma (a_j \emph{\textbf{i}}_j);~ \textbf{A} =
\Sigma (A_j \emph{\textbf{i}}_j);~\nabla = \Sigma
(\emph{\textbf{i}}_j
\partial_j)$.

The gravitational strength $\mathbb{B}_g$ in Eq.(57) includes two
components, $\textbf{g}/c = \partial_0 \emph{\textbf{a}} + \nabla
a_0$ and $\textbf{b} = \nabla \times \emph{\textbf{a}}$ , meanwhile
the electromagnetic strength $\mathbb{B}_e$ involves two parts,
$\textbf{E}/c = ( \partial_0 \emph{\textbf{A}} + \nabla A_0 ) \circ
\emph{\textbf{I}}_0 $ and $\textbf{B} = -( \nabla \times
\emph{\textbf{A}} ) \circ \emph{\textbf{I}}_0$ .

When the coordinates are rotated, we have the octonion strength
$\mathbb{B}' (b'_0, b'_1, b'_2, b'_3, B'_0, B'_1, B'_2, B'_3)$ from
Eq.(45). And then, we have the invariable scalar strength of
gravitational field by Eq.(46).
\begin{eqnarray}
b_0 = b'_0
\end{eqnarray}

The above means that the scalar potential and the scalar strength of
gravitational field are the invariants in the octonion space.

\subsection{Conservation of mass}

The linear momentum density $\mathbb{S}_g = m \mathbb{V}_g $ is the
part source of the gravitational field, and the electric current
density $\mathbb{S}_e = q \mathbb{V}_g \circ \emph{\textbf{I}}_0$ is
that of the electromagnetic field. They combine together to become
the source $\mathbb{S}$ .
\begin{eqnarray}
\mu \mathbb{S} = - ( \mathbb{B}/v_0 + \lozenge)^* \circ \mathbb{B} =
\mu_g^g \mathbb{S}_g + k_{eg} \mu_e^g \mathbb{S}_e - \mathbb{B}^*
\circ \mathbb{B}/v_0
\end{eqnarray}
where, $k_{eg}^2 = \mu_g^g /\mu_e^g$; $q$ is the electric charge
density; $\mu_e^g$ is the constant; $m$ is the inertial mass
density.

In some cases, the electric charge is combined with the mass to
become the electron or proton etc., we have the condition $ R_i
\emph{\textbf{I}}_i = r_i \emph{\textbf{i}}_i \circ
\emph{\textbf{I}}_0$ and $V_i \emph{\textbf{I}}_i = v_i
\emph{\textbf{i}}_i \circ \emph{\textbf{I}}_0$ .

The $\mathbb{B}^* \circ \mathbb{B}/(2\mu_g^g)$ is the energy
density, and includes that of the electromagnetic field.
\begin{eqnarray}
\mathbb{B}^* \circ \mathbb{B}/ \mu_g^g = \mathbb{B}_g^* \circ
\mathbb{B}_g / \mu_g^g + \mathbb{B}_e^* \circ \mathbb{B}_e / \mu_e^g
\end{eqnarray}

The octonion linear momentum density is
\begin{eqnarray}
\mathbb{P} = \mu \mathbb{S} / \mu_g^g = \widehat{m} v_0 + \Sigma (m
v_j \emph{\textbf{i}}_j ) + \Sigma (M V_i \emph{\textbf{i}}_i \circ
\emph{\textbf{I}}_0 )
\end{eqnarray}
where, $\widehat{m} = m - (\mathbb{B}^* \circ \mathbb{B} / \mu_g^g
)/v_0^2 $; $M = k_{eg} \mu_e^g q / \mu_g^g$; $\emph{\textbf{i}}_0 =
1$.

The above means that the gravitational mass density $\widehat{m}$ is
changed with the strength of the electromagnetic field or the
gravitational field.

By Eq.(45), we have the linear momentum density, $\mathbb{P}'
(\widehat{m}' v'_0 , m' v'_1 , m' v'_2 , m' v'_3 , M' V'_0 , M' V'_1
, M' V'_2 , M' V'_3 )$, when the octonion coordinate system is
rotated.
\begin{eqnarray}
\widehat{m}v_0 = \widehat{m}' v'_0
\end{eqnarray}

Under Eqs.(52) and (62), we obtain the conservation of mass as
follows.
\begin{eqnarray}
\widehat{m} = \widehat{m}'
\end{eqnarray}

The above means that if we emphasize the definition of velocity and
Eq.(61), either the inertial mass density or the gravitational mass
density will remain the same, under the octonion transformation in
the electromagnetic field and gravitational field.

\subsection{Mass continuity equation}

In the octonion space, the applied force density $\mathbb{F}$ is
defined from the linear momentum density $\mathbb{P}$ in Eq.(61).
\begin{eqnarray}
\mathbb{F} = v_0 (\mathbb{B}/v_0 + \lozenge )^* \circ \mathbb{P}
\end{eqnarray}
where, the applied force includes the gravity, the inertial force,
the Lorentz force, and the interacting force between the magnetic
strength with magnetic moment, etc.

The applied force density $\mathbb{F}$ is rewritten as follows.
\begin{eqnarray}
\mathbb{F} = f_0 + \Sigma (f_j \emph{\textbf{i}}_j ) + \Sigma (F_i
\emph{\textbf{I}}_i )
\end{eqnarray}
where, $f_0 = \partial p_0 / \partial t + v_0 \Sigma (  \partial p_j
/ \partial r_j ) + \Sigma ( b_j p_j + B_j P_j ) $; $p_0 =
\widehat{m} v_0$, $p_j = m v_j $; $P_i = M V_i $; $\triangle m = - (
\mathbb{B} \circ \mathbb{B}/\mu_g^g)/v_0^2$.

When the coordinate system rotates, we have the octonion applied
force density $\mathbb{F}' (f'_0, f'_1, f'_2, f'_3, F'_0, F'_1,
F'_2, F'_3)$. And then, we have the result by Eq.(46).
\begin{eqnarray}
f_0 = f'_0
\end{eqnarray}

When the right side is zero in the above, we have the mass
continuity equation in the case for coexistence of the gravitational
field and electromagnetic field.
\begin{eqnarray}
\partial \widehat{m} / \partial t + \Sigma ( \partial p_j /
\partial r_j ) + \Sigma ( b_j p_j  + B_j P_j ) / v_0 = 0
\end{eqnarray}

Further, if the strength is zero, $b_j = B_j = 0$, the above will be
reduced to the following equation.
\begin{eqnarray}
\partial m / \partial t + \Sigma ( \partial p_j /
\partial r_j ) = 0
\end{eqnarray}

The above states that the gravitational strength and electromagnetic
strength have the influence on the mass continuity equation,
although the $\Sigma ( b_j p_j +  B_j P_j ) / v_0$ and the
$\triangle m$ both are usually very tiny when the fields are weak.
When we emphasize the definitions of applied force and velocity in
gravitational and electromagnetic fields, the mass continuity
equation will be the invariant equation under the octonion
transformation.

\subsection{Spin angular momentum}

In the octonion space, the angular momentum density is defined from
Eqs.(47) and (61).
\begin{eqnarray}
\mathbb{L} = (\mathbb{R} + k_{rx} \mathbb{X} ) \circ \mathbb{P}
\end{eqnarray}
where, $\mathbb{X} = \Sigma ( x_i \emph{\textbf{i}}_i ) + \Sigma (
X_i \emph{\textbf{I}}_i )$; the angular momentum density includes
the orbital angular momentum density and spin angular momentum
density in the gravitational and electromagnetic fields; $k_{rx} =
1$ is the coefficient.

The angular momentum density is rewritten as
\begin{eqnarray}
\mathbb{L} = l_0 + \Sigma (l_j \emph{\textbf{i}}_j ) + \Sigma (L_i
\emph{\textbf{I}}_i )
\end{eqnarray}
where, the scalar part of angular momentum density
\begin{eqnarray}
l_0 = (r_0 + k_{rx} x_0 ) p_0 - \Sigma \left\{(r_j + k_{rx} x_j )
p_j \right\} - \Sigma \left\{(R_i + k_{rx} X_i ) P_i \right\}
\end{eqnarray}

The $l_0$ is considered as the spin angular momentum density in the
gravitational and electromagnetic fields.

When the coordinate system rotates, we have the octonion angular
momentum density $\mathbb{L}' = \Sigma ( l'_i \emph{\textbf{i}}'_i +
L'_i \emph{\textbf{I}}'_i )$. Under the octonion coordinate
transformation, we have the conservation of spin from Eq.(46).
\begin{eqnarray}
l_0 = l'_0
\end{eqnarray}

The above means that the spin angular momentum density $l_0$ is
invariable in the case for coexistence of the gravitational field
and electromagnetic field, under the octonion transformation.

\begin{table}[b]
\caption{\label{tab:table1}The definitions and their mechanical
invariants in the case for coexistence of the gravitational field
and electromagnetic field in the octonion space.}
\begin{ruledtabular}
\begin{tabular}{llll}
$definition$                             & $invariant $       &    $ meaning $                  & $theorem$          \\
\hline
$\mathbb{R}$                             & $r_0 = r'_0$       &    $invariable~scalar$          & $-$                \\
$\mathbb{V}$                             & $v_0 = v'_0$       &    $invariable~speed~of~light$  & $-$                \\
$\mathbb{X}$                             & $x_0 = x'_0$       &    $invariable~scalar$          & $-$                \\
$\mathbb{A}$                             & $a_0 = a'_0$       &    $invariable~gravitational~scalar~potential$ & $-$ \\
$\mathbb{B}$                             & $b_0 = b'_0$       &    $invariable~gravitational~gauge$            & $-$ \\
$\mathbb{P}$                             & $p_0 = p'_0$       &    $conservation~of~mass$       & $-$                \\
$\mathbb{F}$                             & $f_0 = f'_0$       &    $mass~continuity~equation$   & $linear~momentum$  \\
$\mathbb{L}$                             & $l_0 = l'_0$       &    $invariable~spin~angular~momentum~density$  & $-$ \\
$\mathbb{W}$                             & $w_0 = w'_0$       &    $conservation~of~energy$     & $angular~momentum$ \\
$\mathbb{N}$                             & $n_0 = n'_0$       &    $energy~continuity~equation$ & $torque$           \\
\end{tabular}
\end{ruledtabular}
\end{table}

\subsection{Conservation of energy}

The total energy density $\mathbb{W}$ is defined from the angular
momentum density $\mathbb{L}$ in Eq.(69).
\begin{eqnarray}
\mathbb{W}  = v_0 ( \mathbb{B}/v_0 + \lozenge) \circ \mathbb{L}
\end{eqnarray}
where, the energy includes the electric potential energy, the
gravitational potential energy, the interacting energy between the
magnetic moment with magnetic strength, the torque between the
electric moment with electric strength, and the work, etc. in the
gravitational field and electromagnetic field.

The above can be rewritten as,
\begin{eqnarray}
\mathbb{W} = w_0 + \Sigma (w_j \emph{\textbf{i}}_j ) + \Sigma (W_i
\emph{\textbf{I}}_i )
\end{eqnarray}
where, the scalar $w_0 = v_0 \partial_0 l_0 + (v_0 \nabla +
\emph{\textbf{h}}) \cdot \emph{\textbf{j}} + \emph{\textbf{H}} \cdot
\emph{\textbf{J}} $ ; $\emph{\textbf{h}} = \Sigma (b_j
\emph{\textbf{i}}_j)$; $\emph{\textbf{H}} = \Sigma (B_j
\emph{\textbf{I}}_j)$; $\emph{\textbf{j}} = \Sigma (l_j
\emph{\textbf{i}}_j)$; $\emph{\textbf{J}} = \Sigma (L_j
\emph{\textbf{I}}_j)$.

When the coordinate system rotates, we have the octonion energy
density $\mathbb{W}' = \Sigma ( w'_i \emph{\textbf{i}}'_i + W'_i
\emph{\textbf{I}}'_i )$. Under the octonion transformation, the
scalar part of total energy density is the energy density. And then
we have the conservation of energy as follows.
\begin{eqnarray}
w_0 = w'_0
\end{eqnarray}

In some special cases, the right side is equal to zero. We obtain
the spin continuity equation.
\begin{eqnarray}
\partial l_0 / \partial r_0 + \nabla \cdot \emph{\textbf{j}}
+ (\emph{\textbf{h}} \cdot \emph{\textbf{j}} + \emph{\textbf{H}}
\cdot \emph{\textbf{J}}) / v_0 = 0
\end{eqnarray}

If the last term is neglected, the above is reduced to
\begin{eqnarray}
\partial l_0 / \partial r_0 + \nabla \cdot \emph{\textbf{j}} = 0
\end{eqnarray}
further, if the last term is equal to zero, we have
\begin{eqnarray}
\partial l_0 / \partial t = 0
\end{eqnarray}
where, when the time $t$ is only the independent variable, the
$\partial l_0 / \partial t$ will become the $d l_0 / d t$ .

The above means the energy density $w_0$ is invariable in the case
for coexistence of the gravitational field and the electromagnetic
field with the octonion transformation.

\subsection{Energy continuity equation}

In the octonion space, the external power density $\mathbb{N}$ is
defined from the total energy density in Eq.(73).
\begin{eqnarray}
\mathbb{N}  = v_0 ( \mathbb{B}/v_0 + \lozenge)^* \circ \mathbb{W}
\end{eqnarray}
where, the external power density $\mathbb{N}$ includes the power
density in the gravitational and electromagnetic fields.

The external power density can be rewritten as follows.
\begin{eqnarray}
\mathbb{N} = n_0 + \Sigma (n_j \emph{\textbf{i}}_j ) + \Sigma (N_i
\emph{\textbf{I}}_i )
\end{eqnarray}
where, the scalar $n_0 = v_0 \partial_0 w_0 + (v_0 \nabla +
\emph{\textbf{h}})^* \cdot \emph{\textbf{y}} + \emph{\textbf{H}}^*
\cdot \emph{\textbf{Y}} $; $\emph{\textbf{y}} = \Sigma (w_j
\emph{\textbf{i}}_j)$; $\emph{\textbf{Y}} = \Sigma (W_j
\emph{\textbf{I}}_j)$.

When the coordinate system rotates, we have the octonion external
power density $\mathbb{N}' = \Sigma ( n'_i \emph{\textbf{i}}'_i +
N'_i \emph{\textbf{I}}'_i )$. Under the octonion coordinate
transformation, the scalar part of external power density is the
power density and remains unchanged by Eq.(46).
\begin{eqnarray}
n_0 = n'_0
\end{eqnarray}

In a special case, the right side is equal to zero. And then, we
obtain the energy continuity equation.
\begin{eqnarray}
\partial_0 w_0 + \nabla^* \cdot \emph{\textbf{y}} +
(\emph{\textbf{h}}^* \cdot \emph{\textbf{y}} + \emph{\textbf{H}}^*
\cdot \emph{\textbf{Y}}) / v_0 = 0
\end{eqnarray}

If the last term is neglected, the above is reduced to
\begin{eqnarray}
\partial_0 w_0 + \nabla^* \cdot \emph{\textbf{y}} = 0
\end{eqnarray}
further, if the last term is equal to zero, we have
\begin{eqnarray}
\partial w_0 / \partial t = 0
\end{eqnarray}
where, when the time $t$ is only the independent variable, the
$\partial w_0 / \partial t$ will become the $d w_0 / d t$ .

The above means the power density $n_0$ is the invariant in the case
for coexistence of the gravitational field and electromagnetic
field, under the octonion transformation.

In other words, either the strength or torque density has an
influence on the energy continuity equation in the gravitational
field and the electromagnetic field.

In the octonion space, the deductive results about the conservation
laws and the scalar invariants depend on the definition combinations
in the case for coexistence of gravitational field and
electromagnetic field. By means of the definition combination of the
radius vector and the velocity, we obtain the results of the
invariable speed of light, Galilean transformation, and Lorentz
transformation. Choosing the definition combination of the linear
momentum, the operator $\lozenge$, and the velocity, we have the
conclusions about the conservation of mass and the mass continuity
equation in the gravitational and electromagnetic fields. With the
definition combination of the angular momentum, the velocity, and
the operator $\lozenge$, we obtain the inferences of the
conservation of spin, the spin continuity equation, the conservation
of energy, and the energy continuity equation, etc.

\section{Electric invariants in electromagnetic field}

In the electromagnetic field, there are some special invariants,
which is different to the mechanical invariants. They are called the
electric invariants, which include the conservation of charge, the
charge continuity equation, and the spin magnetic moment etc. The
conservation of charge is similar to the conservation of mass, and
is investigated by a lot of scientists. This conservation law is an
important theorem in the physics, with the numerous applications in
the quantum mechanics, the electromagnetic theory \cite{heaviside},
and the optics theory, etc. With the characteristics of octonion, we
obtain some scalar invariants, conservation laws, and theorems in
the case for coexistence of the electromagnetic field and
gravitational field, under the octonion coordinate transformation.

\subsection{Electromagnetic and gravitational fields}

The invariants of the physical quantities in the gravitational field
and the electromagnetic field can be illustrated by quaternions,
although their definitions will be extended in some aspects.

In the quaternion space for the gravitational field, the radius
vector is $\mathbb{R}_g = \Sigma (r_i \emph{\textbf{i}}_i)$, the
velocity $\mathbb{V}_g = \Sigma (v_i \emph{\textbf{i}}_i)$, the
basis vector $\mathbb{E}_g$ = ($\emph{\textbf{i}}_0$,
$\emph{\textbf{i}}_1$, $\emph{\textbf{i}}_2$,
$\emph{\textbf{i}}_3$), with $\emph{\textbf{i}}_0 = 1$. In the
quaternion space for the electromagnetic field, the radius vector is
$\mathbb{R}_e = \Sigma (R_i \emph{\textbf{I}}_i)$, the velocity
$\mathbb{V}_e = \Sigma (V_i \emph{\textbf{I}}_i)$, the basis vector
$\mathbb{E}_e$ = ($\emph{\textbf{I}}_0$, $\emph{\textbf{I}}_1$,
$\emph{\textbf{I}}_2$, $\emph{\textbf{I}}_3$), with $\mathbb{E}_e$ =
$\mathbb{E}_g$ $\circ$ $\emph{\textbf{I}}_0$ . Two quaternion spaces
combine together to become the octonion space. In the octonion
space, the radius vector $\mathbb{R} = \Sigma (r_i
\emph{\textbf{i}}_i ) + \Sigma ( R_i \emph{\textbf{I}}_i)$, the
velocity $ \mathbb{V} = \Sigma (v_i \emph{\textbf{i}}_i ) + \Sigma
(V_i \emph{\textbf{I}}_i )$, and the basis vector $ \mathbb{E} =
(\emph{\textbf{i}}_0, \emph{\textbf{i}}_1, \emph{\textbf{i}}_2,
\emph{\textbf{i}}_3, \emph{\textbf{I}}_0, \emph{\textbf{I}}_1,
\emph{\textbf{I}}_2, \emph{\textbf{I}}_3) $. The $r_0 = v_0 t$;
$v_0$ is the speed of light; $t$ denotes the time. The $R_0 = V_0
T$; $V_0$ is the speed of light-like; $T$ is a time-like quantity.
The $\circ$ denotes the octonion multiplication. In some special
cases, the electric charge is combined with the mass to become the
electron or the proton etc. And then $ R_i \emph{\textbf{I}}_i = r_i
\emph{\textbf{i}}_i \circ \emph{\textbf{I}}_0$ and $ V_i
\emph{\textbf{I}}_i = v_i \emph{\textbf{i}}_i \circ
\emph{\textbf{I}}_0$ .

In the octonion space, $\mathbb{A}_g = (a_0 , a_1 , a_2 , a_3)$ is
the gravitational potential, and $\mathbb{A}_e = (A_0 , A_1 , A_2 ,
A_3)$ is the electromagnetic potential. The octonion potential
\begin{eqnarray}
\mathbb{A} = \mathbb{A}_g + k_{eg} \mathbb{A}_e
\end{eqnarray}
where, $k_{eg}$ is the coefficient.

The octonion strength $\mathbb{B} = \Sigma (b_i \emph{\textbf{i}}_i
) + \Sigma ( B_i \emph{\textbf{I}}_i)$ includes the gravitational
strength $\mathbb{B}_g$ and the electromagnetic strength
$\mathbb{B}_e$ . And the gauge $b_0 = 0$ and $B_0 = 0$ .
\begin{eqnarray}
\mathbb{B} = \lozenge \circ \mathbb{A} = \mathbb{B}_g + k_{eg}
\mathbb{B}_e
\end{eqnarray}
where, the gravitational strength $\mathbb{B}_g$ includes two parts,
$\textbf{g} = ( g_{01} , g_{02} , g_{03} ) $ and $\textbf{b} = (
g_{23} , g_{31} , g_{12} )$, while the electromagnetic strength
$\mathbb{B}_e$ involves two components, $\textbf{E} = ( B_{01} ,
B_{02} , B_{03} ) $ and $\textbf{B} = ( B_{23} , B_{31} , B_{12} )$
.

The linear momentum density $\mathbb{S}_g = m \mathbb{V}_g $ is the
source for the gravitational field, and the electric current density
$\mathbb{S}_e = q \mathbb{V}_g \circ \emph{\textbf{I}}_0$ for the
electromagnetic field. And they can be combined together to become
the source $\mathbb{S}$ .
\begin{eqnarray}
\mu \mathbb{S} = - ( \mathbb{B}/v_0 + \lozenge)^* \circ \mathbb{B} =
\mu_g^g \mathbb{S}_g + k_{eg} \mu_e^g \mathbb{S}_e - \mathbb{B}^*
\circ \mathbb{B}/v_0
\end{eqnarray}
where, $\mu_e^g$ is the constant; $*$ denotes the conjugate of
octonion. And
\begin{eqnarray}
\mathbb{B}^* \circ \mathbb{B}/ \mu_g^g = \mathbb{B}_g^* \circ
\mathbb{B}_g / \mu_g^g + \mathbb{B}_e^* \circ \mathbb{B}_e / \mu_e^g
\end{eqnarray}

In the octonion space, we have the applied force density,
\begin{eqnarray}
\mathbb{F} =  v_0 ( \mathbb{B}/v_0 + \lozenge)^* \circ \mathbb{P}
\end{eqnarray}
and the angular momentum density
\begin{eqnarray}
\mathbb{L} = (\mathbb{R} + k_{rx} \mathbb{X} ) \circ \mathbb{P}
\end{eqnarray}
where, the linear momentum $\mathbb{P} = \mu \mathbb{S} / \mu_g^g$ ;
the quantity $\mathbb{X} = \Sigma ( x_i \emph{\textbf{i}}_i ) +
\Sigma ( X_i \emph{\textbf{I}}_i )$ .

By means of the total energy density,
\begin{eqnarray}
\mathbb{W}  = v_0 ( \mathbb{B}/v_0 + \lozenge) \circ \mathbb{L}
\end{eqnarray}
we obtain the external power density
\begin{eqnarray}
\mathbb{N}  = v_0 ( \mathbb{B}/v_0 + \lozenge)^* \circ \mathbb{W}
\end{eqnarray}
where, the external power density includes the power density in the
gravitational and electromagnetic fields.

\subsection{Speed of light-like}
The physical quantity $\mathbb{D}(d_0, d_1, d_2, d_3, D_0, D_1, D_2,
D_3)$ in octonion space is defined as
\begin{eqnarray}
\mathbb{D} = d_0 + d_1 \emph{\textbf{i}}_1 + d_2 \emph{\textbf{i}}_2
+ d_3 \emph{\textbf{i}}_3 +  D_0 \emph{\textbf{I}}_0 + D_1
\emph{\textbf{I}}_1 +  D_2 \emph{\textbf{I}}_2 + D_3
\emph{\textbf{I}}_3
\end{eqnarray}

In case of the coordinate system is transformed into another, the
physical quantity $\mathbb{D}$ will be transformed into $\mathbb{D}'
(d'_0 , d'_1 , d'_2 , d'_3 , D'_0 , D'_1 , D'_2 , D'_3 )$ .
\begin{equation}
\mathbb{D}' = \mathbb{K}^* \circ \mathbb{D} \circ \mathbb{K}
\end{equation}
where, $\mathbb{K}$ is the octonion, and $\mathbb{K}^* \circ
\mathbb{K} = 1$.

In the above, the scalar part is one and the same during the
octonion coordinate system is transforming. So
\begin{eqnarray}
d_0 = d'_0
\end{eqnarray}

In the octonion space, the velocity
\begin{eqnarray}
\mathbb{V} =  \Sigma (v_i \emph{\textbf{i}}_i ) + \Sigma (V_i
\emph{\textbf{I}}_i )
\end{eqnarray}
and a new physical quantity $\mathbb{V}_q$ can be defined from the
above,
\begin{eqnarray}
\mathbb{V}_q = \mathbb{V} \circ \emph{\textbf{I}}_0^* = \Sigma (V_i
\emph{\textbf{i}}_i ) - \Sigma (v_i \emph{\textbf{I}}_i )
\end{eqnarray}

By Eq.(94), we have the octonion physical quantity, $\mathbb{V}'_q
(v'_0, v'_1, v'_2, v'_3, V'_0, V'_1, V'_2, V'_3)$, when the
coordinate system is rotated. Under the coordinate transformation,
the scalar part of $\mathbb{V}_q$ remains unchanged.
\begin{eqnarray}
V_0 = V'_0
\end{eqnarray}

The above means that the speed of light-like, $V_0$, will remain the
same in the electromagnetic field and gravitational field during the
octonion coordinate transformation.

\subsection{Conservation of charge}
In the octonion space, the linear momentum density $\mathbb{P}$ is
\begin{eqnarray}
\mathbb{P} = \mu \mathbb{S} / \mu_g^g = \widehat{m} v_0 + \Sigma (m
v_j \emph{\textbf{i}}_j ) + \Sigma (M V_i \emph{\textbf{I}}_i )
\end{eqnarray}
where, the gravitational mass density $\widehat{m} = m + \triangle m
$ ; $ \triangle m = - (\mathbb{B}^* \circ \mathbb{B} / \mu_g^g
)/v_0^2 $ ; $M = k_{eg} \mu_e^g q / \mu_g^g$ . $q$ is the electric
charge density; $m$ is the inertial mass density.

A new physical quantity $\mathbb{P}_q$ can be defined from the
above,
\begin{eqnarray}
\mathbb{P}_q = \mathbb{P} \circ \emph{\textbf{I}}_0^* = M V_0 +
\Sigma (M V_j \emph{\textbf{i}}_j ) - \left\{ \widehat{m} v_0
\emph{\textbf{I}}_0 + \Sigma (m v_j \emph{\textbf{I}}_j ) \right\}
\end{eqnarray}

By Eq.(94), we have the linear momentum density, $\mathbb{P}'
(\widehat{m}' v'_0 , m' v'_1 , m' v'_2 , m' v'_3 , M' V'_0 , M' V'_1
, M' V'_2 , M' V'_3 )$, when the octonion coordinate system is
rotated. Under the coordinate transformation, the scalar part of
$\mathbb{P}_q$ remains unchanged.
\begin{eqnarray}
M V_0 = M' V'_0 \nonumber
\end{eqnarray}

With Eq.(98) and the above, we obtain the conservation of charge as
follows. And $M$ is the scalar invariant, which is in direct
proportion to the electric charge density $q$ .
\begin{eqnarray}
M = M'
\end{eqnarray}

The above means that if we emphasize the definitions of the velocity
and the linear momentum, the electric charge density will remain the
same, under the octonion coordinate transformation in the
electromagnetic field and gravitational field.

\subsection{Charge continuity equation}
In the octonion space, the applied force density
\begin{eqnarray}
\mathbb{F} = f_0 + \Sigma (f_j \emph{\textbf{i}}_j ) + \Sigma (F_i
\emph{\textbf{I}}_i )
\end{eqnarray}
where, $F_0 = v_0 \partial P_0 / \partial r_0 + v_0 \Sigma (
\partial P_j / \partial r_j ) + \Sigma ( b_j P_j - B_j p_j ) $; $p_0
= \widehat{m} v_0$, $p_j = m v_j $; $P_i = M V_i $ .

A new physical quantity $\mathbb{F}_q$ can be defined from the
above,
\begin{eqnarray}
\mathbb{F}_q = \mathbb{F} \circ \emph{\textbf{I}}_0^* =  F_0 +
\Sigma (F_j \emph{\textbf{i}}_j ) - \Sigma (f_i \emph{\textbf{I}}_i
)
\end{eqnarray}

When the coordinate system rotates, we have the octonion applied
force density $\mathbb{F}' (f'_0, f'_1, f'_2, f'_3, F'_0, F'_1,
F'_2, F'_3)$, the radius vector $\mathbb{R}' (r'_0, r'_1, r'_2,
r'_3, R'_0, R'_1, R'_2, R'_3)$, the momentum density $\mathbb{P}'
(p'_0, p'_1, p'_2, p'_3, P'_0, P'_1, P'_2, P'_3)$, and the strength
$\mathbb{B}' (b'_0, b'_1, b'_2, b'_3, B'_0, B'_1, B'_2, B'_3)$, with
the gauge $b'_0 = B'_0 = 0$. Under the coordinate transformation,
the scalar part of $\mathbb{F}_q$ remains unchanged.
\begin{eqnarray}
F_0 = F'_0
\end{eqnarray}

When the right side is equal to zero in the above, we have the
charge continuity equation in the case for coexistence of the
gravitational field and electromagnetic field.
\begin{eqnarray}
\partial P_0 / \partial r_0 + \Sigma ( \partial P_j /
\partial r_j ) + \Sigma ( b_j P_j - B_j p_j ) / v_0 = 0
\end{eqnarray}

Further, if the last term is neglected, the above equation will be
reduced to,
\begin{eqnarray}
\partial M / \partial t + \Sigma ( \partial P_j / \partial r_j ) = 0
\end{eqnarray}

The above states that the gravitational strength and electromagnetic
strength have the small influence on the charge continuity equation,
although the $\Sigma ( b_j P_j -  B_j p_j ) / v_0$ and $\triangle m$
both are usually very tiny when the field is weak. The charge
continuity equation is the invariant under the octonion coordinate
transformation etc, while the definition of linear momentum density
has to be extended in the gravitational field and electromagnetic
field.

Comparing Eq.(65) with Eq.(103), we find that the mass continuity
equation Eq.(68) and charge continuity equation Eq.(106) can't be
effective at the same time. That states that some mechanical
invariants and electrical invariants will not be valid
simultaneously.

\subsection{Conservation of spin magnetic moment}
In the octonion space, the angular momentum density
\begin{eqnarray}
\mathbb{L} = l_0 + \Sigma (l_j \emph{\textbf{i}}_j ) + \Sigma (L_i
\emph{\textbf{I}}_i )
\end{eqnarray}
where, $L_0 = (r_0 + k_{rx} x_0 ) P_0 - \Sigma \left\{(r_j + k_{rx}
x_j ) P_j \right\} + \Sigma \left\{(R_i + k_{rx} X_i ) p_i \right\}
$ .

A new physical quantity $\mathbb{L}_q$ can be defined from the
above,
\begin{eqnarray}
\mathbb{L}_q = \mathbb{L} \circ \emph{\textbf{I}}_0^* = L_0 + \Sigma
(L_j \emph{\textbf{i}}_j ) - \Sigma (l_i \emph{\textbf{I}}_i )
\end{eqnarray}

When the octonion coordinate system rotates, we have the octonion
linear momentum density $\mathbb{P}' = \Sigma ( p'_i
\emph{\textbf{i}}'_i + P'_i \emph{\textbf{I}}'_i )$, the physical
quantity $\mathbb{X}' = \Sigma ( x'_i \emph{\textbf{i}}'_i + X'_i
\emph{\textbf{I}}'_i )$, the radius vector $\mathbb{R}' = \Sigma (
r'_i \emph{\textbf{i}}'_i + R'_i \emph{\textbf{I}}'_i )$, and the
angular momentum density $\mathbb{L}' = \Sigma ( l'_i
\emph{\textbf{i}}'_i + L'_i \emph{\textbf{I}}'_i )$ respectively
from Eq.(94).

Under the coordinate transformation, the scalar part of
$\mathbb{L}_q$ deduces the conservation of spin magnetic moment.
\begin{eqnarray}
L_0 = L'_0
\end{eqnarray}

The above means that the spin magnetic moment density $L_0$ is
invariable in the case for coexistence of the gravitational field
and electromagnetic field, under the octonion coordinate
transformation, etc.

\subsection{Continuity equation of spin magnetic moment}
In the octonion space, the total energy density
\begin{eqnarray}
\mathbb{W} = w_0 + \Sigma (w_j \emph{\textbf{i}}_j ) + \Sigma (W_i
\emph{\textbf{I}}_i )
\end{eqnarray}
where, $W_0 = v_0 \partial L_0 / \partial r_0 + v_0 \Sigma (
\partial L_j / \partial r_j ) + \Sigma ( b_j L_j - B_j l_j ) $.

A new physical quantity $\mathbb{W}_q$ can be defined from the
above,
\begin{eqnarray}
\mathbb{W}_q = \mathbb{W} \circ \emph{\textbf{I}}_0^* = W_0 + \Sigma
(W_j \emph{\textbf{i}}_j ) - \Sigma (w_i \emph{\textbf{I}}_i)
\end{eqnarray}

When the coordinate system rotates, we have the octonion strength
$\mathbb{B}' = \Sigma ( b'_j \emph{\textbf{i}}'_j + B'_j
\emph{\textbf{I}}'_j )$, angular momentum density $\mathbb{L}' =
\Sigma ( l'_i \emph{\textbf{i}}'_i + L'_i \emph{\textbf{I}}'_i )$,
radius vector $\mathbb{R}' = \Sigma ( r'_i \emph{\textbf{i}}'_i +
R'_i \emph{\textbf{I}}'_i )$, and the energy density $\mathbb{W}' =
\Sigma ( w'_i \emph{\textbf{i}}'_i + W'_i \emph{\textbf{I}}'_i )$,
from Eq.(94) respectively. And there are the gauge equations $b'_0 =
0 $ and $B'_0 = 0$ .

Under the coordinate transformation, the scalar part of
$\mathbb{W}_q$ remains unchanged. And then, we have the conservation
of energy-like.
\begin{eqnarray}
W_0 = W'_0
\end{eqnarray}

In some special cases, the right side is equal to zero. We obtain
the continuity equation of spin magnetic moment.
\begin{eqnarray}
\partial L_0 / \partial r_0 + \Sigma (
\partial L_j / \partial r_j ) + \Sigma ( b_j L_j - B_j l_j ) / v_0 = 0
\end{eqnarray}

If the last term is neglected, the above is reduced to
\begin{eqnarray}
\partial L_0 / \partial r_0 + \Sigma (
\partial L_j / \partial r_j ) = 0
\nonumber
\end{eqnarray}
further, if the last term is equal to zero, we have
\begin{eqnarray}
\partial L_0 / \partial t = 0
\end{eqnarray}
where, when the time $t$ is only the independent variable, the
$\partial L_0 / \partial t$ will become the $d L_0 / d t$ .

The above means the energy-like density $W_0$ is invariable in the
case for coexistence of the gravitational field and the
electromagnetic field, under the octonion coordinate transformation,
etc.

\begin{table}[b]
\caption{\label{tab:table1}The definitions and their electric
invariants of the physical quantities in the octonion space.}
\begin{ruledtabular}
\begin{tabular}{llll}
$ definition $                           & $ invariant$       &    $conservation$               & $ theorem$         \\
\hline
$\mathbb{R}\circ\emph{\textbf{I}}_0^*$   & $R_0 = R'_0$       &    $invariable~scalar$          & $-$                \\
$\mathbb{V}\circ\emph{\textbf{I}}_0^*$   & $V_0 = V'_0$       &    $invariable~speed~of~light-like$            & $-$ \\
$\mathbb{X}\circ\emph{\textbf{I}}_0^*$   & $X_0 = X'_0$       &    $invariable~scalar$          & $-$                \\
$\mathbb{A}\circ\emph{\textbf{I}}_0^*$   & $A_0 = A'_0$       &    $invariable~electromagnetic~scalar~potential$ & $-$  \\
$\mathbb{B}\circ\emph{\textbf{I}}_0^*$   & $B_0 = B'_0$       &    $invariable~electromagnetic~gauge$          & $-$ \\
$\mathbb{P}\circ\emph{\textbf{I}}_0^*$   & $P_0 = P'_0$       &    $conservation~of~charge$     & $-$                \\
$\mathbb{F}\circ\emph{\textbf{I}}_0^*$   & $F_0 = F'_0$       &    $charge~continuity~equation$ & $electric~current$ \\
$\mathbb{L}\circ\emph{\textbf{I}}_0^*$   & $L_0 = L'_0$       &    $invariable~spin~magnetic~moment~density$   & $-$ \\
$\mathbb{W}\circ\emph{\textbf{I}}_0^*$   & $W_0 = W'_0$       &    $conservation~of~energy-like$     & $magnetic~moment$ \\
$\mathbb{N}\circ\emph{\textbf{I}}_0^*$   & $N_0 = N'_0$       &    $energy-like~continuity~equation$ & $torque-like$ \\
\end{tabular}
\end{ruledtabular}
\end{table}

\subsection{Energy-like continuity equation}
In the octonion space, the external power density
\begin{eqnarray}
\mathbb{N} = n_0 + \Sigma (n_j \emph{\textbf{i}}_j ) + \Sigma (N_i
\emph{\textbf{I}}_i )
\end{eqnarray}
where, $N_0 = v_0 \partial W_0 / \partial r_0 + v_0 \Sigma (
\partial W_j / \partial r_j ) + \Sigma ( b_j W_j - B_j w_j ) $.

A new physical quantity $\mathbb{N}_q$ can be defined from the
above,
\begin{eqnarray}
\mathbb{N}_q = \mathbb{N} \circ \emph{\textbf{I}}_0^* = N_0 + \Sigma
(N_j \emph{\textbf{i}}_j ) - \Sigma (n_i \emph{\textbf{I}}_i)
\end{eqnarray}

When the coordinate system rotates, we have the octonion angular
momentum density $\mathbb{L}' = \Sigma ( l'_i \emph{\textbf{i}}'_i +
L'_i \emph{\textbf{I}}'_i )$, the radius vector $\mathbb{R}' =
\Sigma ( r'_i \emph{\textbf{i}}'_i + R'_i \emph{\textbf{I}}'_i )$,
the energy density $\mathbb{W}' = \Sigma ( w'_i \emph{\textbf{i}}'_i
+ W'_i \emph{\textbf{I}}'_i )$, the strength $\mathbb{B}' = \Sigma (
b'_j \emph{\textbf{i}}'_j + B'_j \emph{\textbf{I}}'_j )$ from
Eq.(94) respectively. And the gauge $b'_0 = B'_0 = 0$ .

Under the coordinate transformation, the scalar part of
$\mathbb{N}_q$ remains unchanged by the above. And then we obtain
the conservation of power-like as follows.
\begin{eqnarray}
N_0 = N'_0
\end{eqnarray}

In a special case, the right side is equal to zero. And then, we
obtain the continuity equation of energy-like.
\begin{eqnarray}
\partial W_0 / \partial r_0 + \Sigma ( \partial W_j / \partial r_j )
+ \Sigma ( b_j W_j - B_j w_j ) / v_0 = 0
\end{eqnarray}

If the last term is neglected, the above is reduced to
\begin{eqnarray}
\partial W_0 / \partial r_0 + \Sigma ( \partial W_j / \partial r_j ) = 0
\nonumber
\end{eqnarray}
further, if the last term is equal to zero, we have
\begin{eqnarray}
\partial W_0 / \partial t = 0
\end{eqnarray}
where, when the time $t$ is only the independent variable, the
$\partial W_0 / \partial t$ will become the $d W_0 / d t$ .

The above means the power-like density $N_0$ is the invariant in the
case for coexistence of the gravitational field and electromagnetic
field, under the Galilean transformation or the Lorentz
transformation, etc.

\section{CONCLUSIONS}

In the quaternion spaces, choosing the definitions of the
$\mathbb{R}$, $\mathbb{V}$, $\mathbb{A}$, $\mathbb{B}$,
$\mathbb{P}$, $\mathbb{X}$, $\mathbb{L}$, $\mathbb{W}$, and
$\mathbb{N}$ in the electromagnetic field, we obtain the
characteristics of some invariants, including the charge continuity
equation and conservation of charge. The charge density and spin
magnetic moment density etc. are the invariants under the octonion
coordinate transformation. The conclusions in the electromagnetic
field can be spread from the quaternion space to the octonion space.
And the definitions of the mass continuity equation etc. will be
extended in the octonion space.

It should be noted that the study for the conservation of mass and
conservation of charge etc. examined only some simple cases under
the octonion coordinate transformation. Despite its preliminary
characteristics, this study can clearly indicate the mass density,
charge density, and spin density etc. are the scalar invariants
respectively, and they are only some simple inferences due to the
weak strength of the electromagnetic field. For the future studies,
the investigation will concentrate on only some predictions about
the mass continuity equation, charge continuity equation, and energy
continuity equation, etc. under the strong potential and strength of
electromagnetic field.

\begin{acknowledgments}
This project was supported partially by the National Natural Science
Foundation of China under grant No.60677039.
\end{acknowledgments}

\end{document}